\documentclass[aps,twocolumn,amsmath,amssymb,showpacs,prl]{revtex4}
\usepackage{epsf}

\newcommand{\etal}{{\it et al.}}

\begin{document}

\title{Hall Effect in Nested Antiferromagnets Near the Quantum Critical
Point}
\author{M. R. Norman$^{1}$, Qimiao Si$^{2}$, Ya. B. Bazaliy$^{1}$,
and R. Ramazashvili$^{1}$}
\affiliation{
$^1$Materials Science Division, Argonne National Laboratory, Argonne,
Illinois 60439\\
$^2$Department of Physics and Astronomy, Rice University, Houston,
Texas 77005-1892}
\date{\today}
\begin{abstract}
We investigate the behavior of the Hall coefficient in the case of
antiferromagnetism driven by Fermi surface nesting, and find that the Hall
coefficient should abruptly increase with the onset of magnetism, as recently
observed in vanadium doped chromium.
This effect is due to the sudden removal of
flat portions of the Fermi surface upon magnetic ordering.  Within this picture,
the Hall coefficient should scale as the square of the
residual resistivity divided by the impurity concentration, which is
consistent with available data.
\end{abstract}

\maketitle

There has been recent interest, both experimentally\cite{GREG,SCHRODER}
and theoretically\cite{JPCM,LQCP}, in the physics of magnetic quantum
critical points (QCPs).  Such QCPs occur when the ordering temperature of a
magnet has been driven to zero continuously by some tuning parameter, such as
chemical doping. Much of the interest is due to strong signatures of
non-Fermi liquid behavior near QCPs, and the difficulties
of standard theories of itinerant magnetism in explaining such behavior.
In the case of heavy fermion metals, there
is some indication that the QCP
is accompanied by localization
of the f electrons\cite{GREG,SCHRODER,JPCM,LQCP}.
This should result in a volume
change of the Fermi surface.
A novel signature
of non-trivial 
QCPs
is a jump in the Hall
coefficient \cite{JPCM,LQCP}.
In the case of heavy fermions, results are  
still preliminary 
at this time\cite{PASCHEN}.
Related issues are also being discussed for high temperature 
superconductors\cite{Chakravarty,Balakirev}.

Recently, Yeh \etal \cite{YEH} studied the Hall coefficient
in the simpler case of
V-doped Cr. Upon doping with V, the N\'{e}el temperature is rapidly
suppressed to zero, leading to a QCP at about 4\% doping.
The Hall coefficient
decreases, 
quite abruptly, by about a factor of two with
doping into the paramagnetic phase. Both the magnitude and the abruptness
of this change are
surprising, given that Cr is well established to be a simple 
spin-density-wave magnet. One of the characteristic features of Cr
is that its Fermi surfaces are nested. Indeed, magnetism in Cr is
traditionally understood as being driven by nesting.

In this paper, we show that the magnitude of the change in the 
zero-temperature Hall coefficient
across the magnetic QCP can be 
quantitatively accounted for by the removal of the flat
portions of the Fermi surface upon magnetic ordering.
From this picture, it follows that the zero-temperature  
Hall coefficient
should scale as
the square of the residual resistivity divided by the impurity
concentration, and we demonstrate that the available data in V-doped
Cr are consistent with such a
relationship. Our results establish that the Fermi-surface nesting
picture is quantitatively correct for zero-temperature properties,
thereby providing a solid foundation for further understanding of
the non-Fermi liquid behavior in this benchmark quantum critical metal.

The Hall conductivity
is in general a component of a tensor object.  In the case
of cubic materials, such as Cr, only one component is relevant. In this paper,
we will confine our discussion to the level of Boltzmann approximation, 
which should be adequate for the zero-temperature limit even when interactions
are 
significant.
Within this approximation, the Hall 
coefficient 
is
\begin{equation}
R_H = \sigma_{xyz}/\sigma_{xx}^2
\label{rh}
\end{equation}
where\cite{PA1},
\begin{equation}
\sigma_{xyz} = \frac{e^3\tau^2}{\hbar\Omega c}\sum_{\vec{k}}
v_x (\vec{v} \times \vec{\nabla}_{k})_z v_y
(-\frac{\partial f}{\partial \epsilon_k})
\label{sigmaxyz}
\end{equation}
\begin{equation}
\sigma_{xx} = \frac{e^2\tau}{\Omega}\sum_{\vec{k}}v_x^2
(-\frac{\partial f}{\partial \epsilon_k})
\label{sigmaxx}
\end{equation}
Here, 1/$\tau$ is the scattering rate,
$\Omega$ the
volume, and $f$ the Fermi distribution function.

The physical picture we propose is as follows.
It is known that parts of the Cr Fermi surface are flat and nested,
and the remaining parts are regular. The magnetic ordering gaps out the flat
surfaces.
We note that Eq.~(\ref{sigmaxyz}) is a weighted sum of various components
of the
inverse mass tensor; the latter measures the curvature of the Fermi surface.
Therefore, the flat Fermi surface sheets will make a small contribution to
$\sigma_{xyz}$ even in the paramagnetic state.
Thus, upon magnetic ordering,
$\sigma_{xyz}$ is not expected to change much.
On the other hand, $\sigma_{xx}$ involves only components of the
velocity 
and would contain considerable contributions from the flat Fermi surface
sheets in the paramagnetic phase. 
Removal of the
flat Fermi surfaces upon magnetic ordering 
should lead to a large change in $\sigma_{xx}$.
This will be
amplified in the Hall
coefficient,
since the square of $\sigma_{xx}$ appears in
Eq.~(\ref{rh}). Such reasoning 
makes clear
a very general relationship between the
the Hall coefficient 
and the longitudinal resistivity across the QCP.

\begin{figure}
\centerline{\epsfxsize=2.4in{\epsfbox{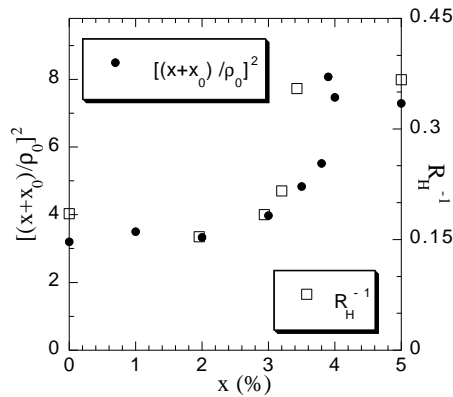}}}
\caption{
Doping dependence of the Hall number ($R_H^{-1}$, carriers per unit cell)
\protect\cite{YEH}
and of the inverse square of the residual resistivity  ($\rho_0$, $\mu \Omega$ cm)
 \protect\cite{JAP}.
$\rho_0$
is divided by
the effective
impurity concentration, $x+x_0$, where $x$ is the doping,
and 
$x_0$ = 0.59\% the
impurity concentration at stochiometry, determined from a linear
fit of
$\rho_0$ at low doping. The discrepancy between the plots is largely
due to the different critical concentrations of the two sample sets
(3.5\% for the Hall samples, 4\% for the resistivity samples).
The experimental $R_H$ point at 5\% doping actually corresponds 
to 10\% doping, and is shown simply to
illustrate the approximate constancy of $R_H$ in the paramagnetic phase.}
\label{rhvsrho0}
\end{figure}

At $T=0$, the inverse of $\sigma_{xx}$ is the residual resistivity, $\rho_0$.
Therefore, we expect scaling between $R_H$ and $\rho_0$. If the tuning
parameter
(such as pressure) does not change the elastic scattering, it follows that
\begin{equation}
\frac{R_H^{pm}}{R_H^{afm}}
=
\left(\frac{\rho_0^{pm}}{\rho_0^{afm}}\right )^2
\label{rh-rho0-pressure}
\end{equation}
where the superscripts ``pm'' and ``afm'' refer to the paramagnetic and
antiferromagntic phases, respectively.
When the transition is induced by doping, the elastic scattering is also
being changed,
presumably in a linear fashion. For impurities that are intermediate
between Born
and unitarity limits, we find that
\begin{equation}
\frac{R_H^{pm}}{R_H^{afm}}
=
\left(\frac{d \rho_0^{pm} / d x}
{d \rho_0^{afm}/dx}
\right )^2
\label{rh-rho0-doping}
\end{equation}
From data 
available 
in the literature\cite{JAP}, the concentration dependence
of $\rho_0$ goes as $x+x_0$, where $x_0$ represents 
impurities already present
in the stochiometric material. We can check the validity of 
Eq.~(\ref{rh-rho0-doping}) by
comparing $(x+x_0)/\rho_0$ 
with the Hall number,
$R_H^{-1}$.
As seen in Fig.~\ref{rhvsrho0}, the correlation is quite good. 
The discrepancy is
mainly due to the fact that the the resistivity was measured on a sample with
critical point at about 4\% doping, whereas the Hall data were taken 
on samples with
the critical doping of 3.5\%. It would be important to test
Eq.~(\ref{rh-rho0-doping}) in greater detail by taking Hall and
resistivity data on
the same samples. We also note that Eq.~(\ref{rh-rho0-pressure})
is consistent with
the pressure data \cite{TOM}.

\begin{figure}
\centerline{\epsfxsize=2.4in{\epsfbox{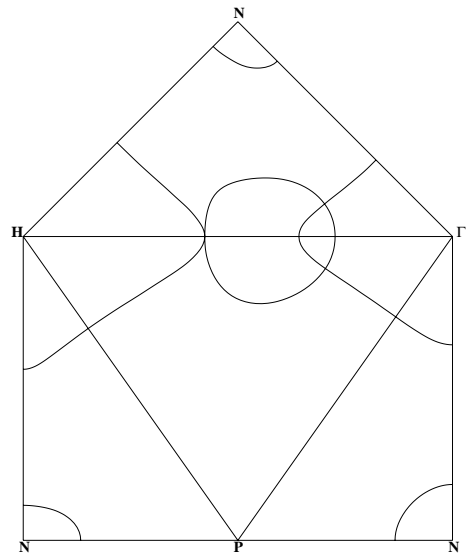}}}
\caption{
Fermi surface of Cr plotted in the faces of the irreducible wedge of
the BCC zone.  In units of $\pi/a$, the symmetry points correspond to
$\Gamma$ (0,0,0), $H$ (2,0,0), $N$ (1,1,0), and $P$ (1,1,1).  The Fermi
surface consists of a $\Gamma$ centered electron octahedron, an $H$
centered hole octahedron, $N$ centered hole ellipsoids, and $\Gamma-H$
centered electron balls.}
\label{bands}
\end{figure}

We now turn to quantitative considerations of the Hall 
coefficient 
itself. To this
effect, we have performed band calculations within the local density
approximation
for Cr, using the linear muffin tin orbital method. After self-consistent
convergence, eigenvalues were generated on a 506 k point grid in the
irreducible
wedge (1/48th) of the BCC Brillouin zone. These eigenvalues were then
interpolated
using a 910 function Fourier series (spline fit).  The resulting Fermi
surface is
shown in Fig.~\ref{bands}. It consists of four parts, a $\Gamma$ centered
electron
octahedron, an H centered hole octahedron, $\Gamma-H$ centered electron
balls, and
$N$ centered hole ellipsoids.  As is well known, the two octahedron
surfaces match up when translated by the magnetic ${\bf Q}$ vector.
The ``gapping out'' of these two surfaces by the magnetic ordering has 
been recently
observed by photoemission\cite{ARPES}.

We show a Fermi surface decomposition of Eqs.~(\ref{rh}-\ref{sigmaxx})
in Table 1.
We see that although the two flat surfaces make up 40\% of the density
of states,
and 52\% of $\sigma_{xx}$, they only make up 22\% of $\sigma_{xyz}$.
The total 
Hall number 
corresponds to 0.54 (in units of carrier concentration),
somewhat larger than the paramagnetic value of 0.37 found for 10\% V
doping\cite{YEH}.  Actually, the theoretical value decreases with V doping
(simulated by a rigid band adjustment of the Fermi energy), and has 
a value of 0.47
for 10\% hole doping.

\begin{table}
\begin{center}
\caption{Decomposition of the transport integrals for
undoped Cr in the paramagnetic phase.  DOS is the density of states,
$\sigma_{xx}$ and $\sigma_{xyz}$ are defined in Eqs.~1-3.
$N-ell$ are the N centered hole
ellipsoids, $H-octa$ the $H$ centered hole octahedron, $\Gamma-octa$ the
$\Gamma$ centered electron octahedron, and $\Gamma-ball$ the $\Gamma-H$
centered electron balls.
Flat 
is the sum of the two octahedra, non-flat
the sum of the rest.  Values listed are the fraction of the total.
The resulting 
Hall number is 0.54, which would be
0.16 if the flat surfaces are eliminated.}
\begin{tabular}{c|cccc|cc}
\hline
\hline
& $N-ell$ & $H-octa$ & $\Gamma-octa$ & $\Gamma-ball$
& flat & non-flat \\
\hline
DOS & 0.12 & 0.19 & 0.21 & 0.48 & 0.40 & 0.60 \\
$\sigma_{xyz}$ & 1.06 & 0.36 & -0.14 & -0.29 & 0.22 & 0.78 \\
$\sigma_{xx}$ & 0.27 & 0.36 & 0.16 & 0.21 & 0.52 & 0.48 \\
\hline
\hline
\end{tabular}
\end{center}
\end{table}

\begin{figure}
\centerline{\epsfxsize=3.4in{\epsfbox{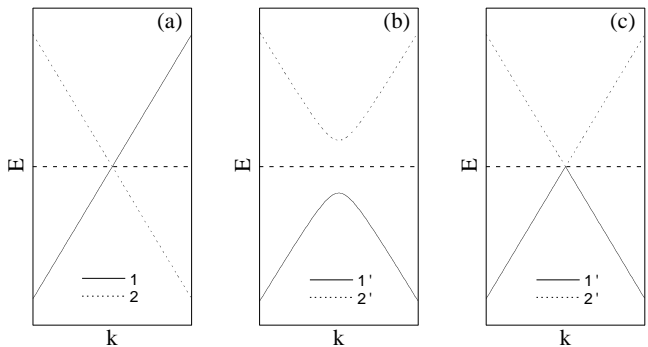}}}
\caption{
Schematic of energy bands in Cr.
(a) paramagnetic phase - band 1 represents the electron octahedron,
band 2 the hole octahedron translated by the magnetic wavevector, ${\bf Q}$.
(b) antiferromagnetic phase - The two bands mix, forming new energy
bands $1^{\prime}$ and $2^{\prime}$ with an energy gap.
(c) limit of (b) as the energy gap
is taken to zero.  Note the different band indexing in (c) as compared to
(a).
The implication of this for transport integrals is discussed in the text.}
\label{crossing}
\end{figure}

If the two octahedron surfaces were completely flat, they would be
immediately removed by magnetic ordering. The Hall number
would then jump from 0.54 to
0.16. (The latter value is identical to experimental values in the magnetic
phase\cite{YEH}.) This can be seen from Fig.~\ref{crossing}.
In Fig.~\ref{crossing}a, we show a schematic electronic
dispersion corresponding to the electron octahedron band of Cr.
Also shown in the schematic is the dispersion of
the hole octahedron band translated by the magnetic wavevector ${\bf Q}$.
The ${\bf Q}$ vector needed for the crossing point to be degenerate with the
Fermi energy will depend on doping (the wavevector is predicted
to be commensurate for electron doping, and increasingly incommensurate with
hole doping, as observed experimentally\cite{RMP}).
Upon magnetic ordering, an energy gap will open up between these two
bands.  If the Fermi energy lies inside the gap, as shown in
Fig.~\ref{crossing}b,
then the contribution of the two bands to
the transport integral is removed.
For the perfectly flat case, this
happens even if the energy gap goes to zero, as in Fig.~\ref{crossing}c.
That is, in this case, a discontinuity in the Hall number is predicted
at the QCP.

In the real band structure, the Fermi surface is not perfectly
flat, and so the crossing point shown in Fig.~\ref{crossing}
disperses as
a function of Fermi surface position.
To illustrate this, we have performed numerical calculations of the
Hall coefficient versus doping.  The most natural way to do this is
by restriction to
the magnetic Brillouin zone.
However, two problems arise.
First, energy gaps will always be present in
such calculations because of the finite grid in $k$ space (this is
easily understood from Fig.~\ref{crossing}).
Second, because of the incommensurability, the zone
can be ill defined.
Instead, we assume a 2 by 2 secular
matrix whose diagonal elements are the eigenvalues of the H centered
octrahedron (translated by ${\bf Q}$), and the electron octahedron, and
whose off-diagonal elements are some constant, $\Delta$ \cite{FOOT}.  Note that 
in this approximation, the $\Gamma-H$ centered electron balls 
and $N$ centered hole ellipsoids are unaffected by magnetic
ordering.
The paramagnetic
electron structure is assumed to be that at 4\% V doping,
so that the only doping dependence is given by the magnetism.
The latter is represented by a ${\bf Q}$ vector of 0.909(2$\pi$/a,0,0),
obtained from the maximum in the susceptibility gotten from
the paramagnetic band eigenvalues (this value agrees with
experiment\cite{RMP}). The gap $\Delta$ (milli-Rydberg) is assumed \cite{RMP}
to vary as 4.9 - 1.3 $x$ (where $x$ is the hole doping in percent).
The results are averaged over the three different ${\bf Q}$ domains.

\begin{figure}
\centerline{\epsfxsize=2.4in{\epsfbox{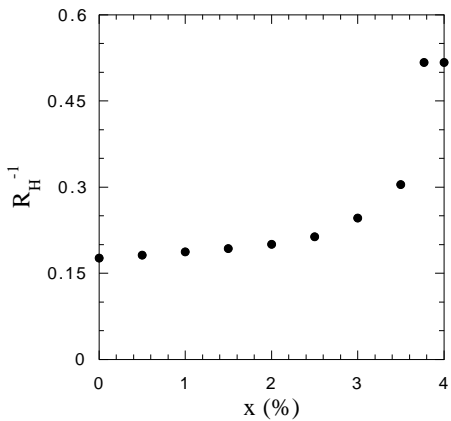}}}
\caption{Calculated 
Hall number as a function of hole doping.}
\label{hallvsx}
\end{figure}

In Fig.~\ref{hallvsx}, we plot the calculated  
Hall number 
as a function of $x$.  Note the striking similarity 
to Fig.~\ref{rhvsrho0}, in particular
the abrupt
drop in the 
Hall number near the QCP (we cannot calculate too
close to the QCP
because of numerical problems which can be understood from
Fig.~\ref{crossing}c).
Although the paramagnetic value is somewhat high, the value in the magnetic
phase is quite close to experiment \cite{YEH}.

The two octahedra are quite flat.
However, on a finer scale, they have different curvatures, leading to 
a continuous
change of the Hall number near the QCP. This change can be
expanded in small
gap $\Delta$.
The coefficients of the expansion are model dependent, but the
leading power in $\Delta$ is universal, and can be easily derived for 
the spherical case using Eqs.~1-3.
For unequal sized spheres which intersect (appropriate for hole 
doped Cr),
$\delta\sigma_{xx},\delta\sigma_{xyz} \sim \Delta$ for any direction
of current, J, and field, B, from which $\delta R_H \sim \Delta$.
For touching spheres, though,
$\delta\sigma_{xx} \sim \Delta$ for $J \parallel Q$,
$\delta\sigma_{xx} \sim \Delta^2$ for $J \perp Q$,
$\delta\sigma_{xyz} \sim \Delta^2$ for $B \parallel Q$,
and $\delta\sigma_{xyz} \sim \Delta$ for B $\perp Q$,
from which $\delta R_H \sim \Delta$ for $B \perp Q$, $J \perp B$
and
$\delta R_H \sim \Delta^2$ for $B \parallel Q$, $J \perp Q$.
(With domain averaging, all changes would go as $\sim \Delta$.)
This touching case should not be 
relevant, though, since it corresponds to where the susceptibility 
has an inflection point as opposed to a maximum \cite{RICE}.
Our results for $\sigma_{xx}$ agree with previous results in the case 
of equal sized spheres \cite{ELLIOTT}.
(In the 2D case, similar conclusions for $R_H$ have been reached in
Ref.~\onlinecite{Chakravarty}.) Since \cite{RMP} 
$\Delta \sim M_{af} \sim x_c-x$ (where $M_{af}$ is 
the antiferromagnetic order parameter) near the
critical concentration $x_c$, $\delta R_H \sim x_c-x$.
(Mean field theory would predict 
$\Delta \sim \sqrt{x_c-x}$, in which case 
$\delta R_H \sim \sqrt{x_c-x}$.)
We note that the
numerical results of Fig.~\ref{hallvsx} are consistent with a much more
rapid variation ($\delta R_H \sim (x_c-x)^{1/4}$), indicating a significant
deviation from the spherical model.

Although the explanation we give for the  behavior of the 
Hall coefficient seems conventional, the result is consistent with the
more exotic physics discussed in Refs.~\onlinecite{JPCM,LQCP}.
In both cases, the Hall
coefficient
jumps because of the
expectation that the Fermi surface 
volume changes
abruptly at the QCP.  In the current case, this is due to nesting.
Presumably, in the heavy fermion case, it is due to disconnection of 
the f electrons
from the Fermi surface.  Still, the net result
for the zero-temperature Hall coefficient
 is the same.
Since entire
regions of the Fermi surface are involved in the phase transition,
the nesting QCP differs
significantly from the standard SDW scenario, where only hot lines of the
Fermi surface are relevant.  As is now well appreciated, hot lines should
not be enough to destabilize the Fermi
liquid \cite{HlubinaRice}, but if entire regions of the
Fermi surface are involved (such as with nesting), the physics changes
considerably.

In fact, nesting may be playing a larger role in QCPs than
has been appreciated.  A recent example is the bilayer ruthenate,
$Sr_3Ru_2O_7$.  This metal exhibits a metamagnetic QCP
accompanied by non-Fermi liquid behavior\cite{SRO}.  Recent neutron
scattering
data find two sets of incommensurate spots, which can be related to
Fermi surface nesting\cite{LUCIA}.
So, it is quite possible that nesting is playing a key role
in this system, and perhaps in heavy fermion metals as well.

In conclusion, we have demonstrated that the abrupt change in
the zero-temperature  
Hall coefficient
in Cr with V doping as observed by Yeh \etal~
can be understood as a
consequence of nesting driven magnetism.  We are able to quantitatively
explain the $T=0$ experimental data by use of band theoretical results,
and have suggested a correlation between the Hall number and the
residual resistivity.
The quantitative success of the Fermi-surface nesting picture should provide
a solid foundation for the eventual understanding of the finite temperature
properties of V-doped Cr, which we do not treat in the current
paper. More generally,
V-doped Cr represents the first
known example of a nesting driven QCP, and further
studies of this system
can shed considerable new light on the 
more exotic QCPs,
such as those seen in heavy fermion metals.

We would like to thank the authors of Refs.~\onlinecite{YEH,TOM}, particularly
Tom Rosenbaum and Anke Husmann, 
for communicating their work to us prior to publication, and permission to
use their data in this paper.
We also thank Dale Koelling for initial help with the band 
calculations, and Lucia Capogna for discussions concerning
neutron data in $Sr_3Ru_2O_7$ and Catherine Pepin concerning pseudogaps.
This work was supported by the U. S. Dept. of Energy, Office of Science,
under Contract No. W-31-109-ENG-38 (MRN, YB, RR) and
by NSF Grant No.\ DMR-0090071, Robert A. Welch Foundation, and TCSAM (QS).
QS acknowledges the hospitality and support
of Argonne National Laboratory,
University of Chicago, University of Illinois
at Urbana-Champaign, and KITP-UCSB.


\begin{thebibliography}{99}

\bibitem{GREG}
G. R. Stewart, Rev. Mod. Phys. {\bf 73}, 797 (2001).

\bibitem{SCHRODER} A.\ Schr\"{o}der {\it et al.},
Nature {\bf 407}, 351 (2000).

\bibitem{JPCM}
P. Coleman, C. Pepin, Q. Si, and R. Ramazashvili, J. Phys. Cond. Matter
{\bf 13}, R723 (2001).

\bibitem{LQCP}
Q.\ Si, S.\ Rabello, K.\ Ingersent, and J.\ L.\ Smith,
Nature {\bf 413}, 804 (2001); cond-mat/0202414 (2002).

\bibitem{PASCHEN}
S. Paschen {\it et al.}, SCES2002 proceedings (2002).

\bibitem{Chakravarty} S. Chakravarty, C. Nayak, S. Tewari, and 
X. Yang, Phys. Rev. Lett. {\bf 89}, 277003 (2002).

\bibitem{Balakirev} F. F. Balakirev, J. B. Betts, A. Migliori,
S. Ono, Y. Ando, and G. B. Boebinger, preprint (2002).

\bibitem{YEH}
A. Yeh, Y.-A. Soh, J. Brooke, G. Aeppli, T. F. Rosenbaum, S. M. Hayden,
Nature {\bf 419}, 459 (2002).

\bibitem{PA1}
J. M. Ziman, {\it Electrons and Phonons} (Oxford Univ. Pr., London,
1960), p. 502-503;
T. P. Beaulac, F. J. Pinski, and P. B. Allen, Phys. Rev. B {\bf 23}, 3617
(1981); W. W. Schulz, P. B. Allen, and N. Trivedi, {\it ibid} {\bf 45},
10886 (1992).

\bibitem{JAP}
J. Takeuchi, H. Sasakura, and Y. Masuda, J. Phys. Soc. Japan {\bf 49},
508 (1980).

\bibitem{TOM}
M. Lee, A. Husmann, T.F. Rosenbaum, and G. Aeppli, to be published
(2002).

\bibitem{ARPES}
J. Schafer \etal, Phys. Rev. Lett. {\bf 83}, 2069 (1999).

\bibitem{RMP}
E. Fawcett, Rev. Mod. Phys. {\bf 60}, 209 (1988); E. Fawcett, H. L. Alberts,
V. Yu. Galkin, D. R. Noakes, J. V. Yakhmi, {\it ibid} {\bf 66}, 25 (1994).

\bibitem{FOOT}
The 2 by 2 secular matrix is exact for a helical SDW.  This is not the
case for the linear SDW, but one can show that in the case considered here,
the linear result for $\sigma_{xx}$, etc., is approximately 
twice the helical result minus the paramagnetic result.

\bibitem{RICE}
T. M. Rice, Phys. Rev. B {\bf 2}, 3619 (1970).

\bibitem{ELLIOTT}
R.J. Eliott and F.A. Wedgwood, Proc. Phys. Soc. {\bf 81}, 846 (1963).

\bibitem{HlubinaRice} R. Hlubina and T. M. Rice,
Phys. Rev. B {\bf 51}, 9253 (1995).

\bibitem{SRO}
R. S. Perry \etal, Phys. Rev. Lett. {\bf 86}, 2661 (2001).

\bibitem{LUCIA}
L. Capogna \etal, Phys. Rev. B {\bf 67}, 012504 (2003).

\end{thebibliography}
\end{document}